\documentclass[preprint,aps,floatfix,superscriptaddress,pre,showpacs]{revtex4}
\usepackage{amsmath, amsthm, amssymb, graphicx}
\input epsf.sty
\begin{document}

\def\g{\gamma}
\def\r{\rho}
\def\w{\omega}
\def\wo{\w_0}
\def\wp{\w_+}
\def\wm{\w_-}
\def\t{\tau}
\def\av#1{\langle#1\rangle}
\def\pf{P_{\rm F}}
\def\pr{P_{\rm R}}
\def\F#1{{\cal F}\left[#1\right]}

\title{Relaxation processes in a system with logarithmic growth}

\author{Mark O. Brown, Robert H. Galyean, Xiangwen Wang, and Michel Pleimling}
\affiliation{Department of Physics, Virginia Tech, Blacksburg, Virginia 24061-0435, USA}

\date{\today}

\begin{abstract}
We discuss relaxation and aging processes in the one- and two-dimensional $ABC$ models. In these driven
diffusive systems of three particle types, biased exchanges in one direction yield a coarsening process characterized
in the long time limit
by a logarithmic growth of ordered domains that take the form of stripes. From the time-dependent length,
derived from the equal-time spatial correlator, and from the mean displacement of individual
particles different regimes in the formation and growth of these domains can be identified.
Analysis of two-times correlation and response functions reveals dynamical scaling in the asymptotic
logarithmic growth regime as well as complicated finite-time and finite-size effects in the early and
intermediate time regimes.
\end{abstract}

\pacs{05.70.Ln,64.60.De,05.10.Ln}

\maketitle

\section{Introduction}

Emerging patterns due to coarsening are encountered in a huge variety of systems \cite{Cro93}, ranging from
magnetic systems \cite{Bra94,Hen10} to complex fluids \cite{Cat15} and from competition in
bacterial colonies \cite{Szo14,Rom13}
to opinion dynamics in social systems \cite{Cas09}.
The general understanding of phase ordering kinetics \cite{Pur09} is well developed for a range of systems, notably
those that are curvature driven and where the size of a typical domain increases with the square root
of time. Power-law growth is also observed in other instances, as for example in systems with a conserved
order parameter undergoing Ostwald ripening \cite{Voo85}. 

However, coarsening processes in more complex systems can yield a domain growth much slower than the
algebraic growth observed in the simpler situations \cite{Cug14}. 
Systems where a logarithmic growth of the form $\sim (\ln t)^{1/\psi}$
is expected include spin glasses and disordered ferromagnets. In general,
it is challenging, both in experimental and numerical studies, to reach the very long timescales needed
to fully access the asymptotic regime. This has been a vexing issue of many years for spin glasses.
On the other hand, disordered ferromagnets, while still being challenging systems, are more easily dealt with
in that regard, and over the years a range of studies focusing on elastic lines in 
disordered media \cite{Kol05,Noh09,Igu09,Mon09}
and on disordered Ising systems \cite{Rao93,Aro08,Par10,Cor11,Cor12,Par12,Cor13,Man14},
including the one-dimensional random field Ising model \cite{Fis01,Cor02},
have shown that the asymptotic growth regime is
indeed non-algebraic and compatible with a logarithmic growth of the domains.

A logarithmically growing length scale is not restricted to disordered systems, but is also encountered 
in some systems dominated by dynamical constraints rather than by quenched disorder \cite{Eva02}. 
Anomalous coarsening with logarithmic domain growth has been identified in a variety of situations with 
dynamical constraints,
as for example in the $ABC$ model introduced by Evans et al \cite{Eva98a,Eva98b}, 
where three different particle types swap places 
asymmetrically, in the model discussed by Lahiri and Ramaswamy \cite{Lah97,Lah00}, where two
sublattices are considered with two types of particles on each sublattice, or in the driven two-lane
particle system studied by Lipowski and Lipowska \cite{Lip09}.
Among these models, the one-dimensional $ABC$ model has enjoyed much attention in recent years
\cite{Cli03,Bod08,Ayy09,Led10a,Led10b,Bar11a,Ber11,Bar11b,Bod11,Coh11,Ger11,Coh12a,Ger12,
Coh12b,Afz13,Ber13,Coh14,Mis14},
due to its unique properties combined with its amenability to exact calculations in some special cases.
Thus the $ABC$ model provides an opportunity to study exactly the long-range correlations
in a system undergoing a non-equilibrium phase transition as well as to elucidate ensemble inequivalence
far from equilibrium.

Whereas most of the papers on the $ABC$ model focused on steady-state properties, only few studies
directly addressed the coarsening process, the logarithmic growth and other issues 
related to this anomalous slow process. After having
identified the anomalous slow domain growth in the $ABC$ model, Evans et al \cite{Eva98b} proposed a simplified 
interface model where domains rather than individual sites are updated. In \cite{Kaf00} slow
coarsening was shown to also exist in a two-dimensional version of the $ABC$ model. Finally, in \cite{Afz13}
the focus was on the interface model in one dimension whose study provided some insights into
aging and dynamical scaling in presence of a logarithmically growing length.

In this paper we aim at further elucidating slow coarsening and relaxation processes in both the
one- and two-dimensional $ABC$ models. In order to do so, we study a range of different quantities 
(the time-dependent space-time correlation function and the length scale obtained from it, two-times
quantities like the autocorrelation and response functions, as well as the mean displacement of
individual particles during the coarsening process) that
allow us to gain a good understanding of the non-equilibrium processes taking place in this system.
We thereby discuss different regimes that can be identified during the coarsening process.
The effects of the system size and of the swapping rate on the domain ordering are discussed.

The paper is organized in the following way. In the next section we recall the model, both in one and
two space dimensions, and introduce the quantities that we study in the remainder of the paper.
Section III is devoted to a discussion of the time-dependent lengths that govern the coarsening process.
Whereas asymptotically the length in the horizontal direction
increases as a logarithm of time, complicated early time regimes emerge,
depending on the degree of asymmetry and the extent of the system. Additional insights are provided by
the mean displacement of individual particles. In section IV we study two-times quantities. Whereas
the autocorrelation exhibits obvious dynamical scaling properties, a complicated response of the system to
perturbations (realized through a sudden change of the swapping rates) is observed. 
We summarize our results in the final section.

\section{Model and quantities}
We consider particles of three different types (called $A$, $B$, and $C$) moving
on a two-dimensional lattice with $L \times M$ sites and periodic boundary conditions in both
the $x$- and $y$-direction. Whereas in the $y$-direction all exchanges are symmetric and happen
with rate 1, in the $x$-direction a bias is introduced that yields an asymmetry in the
exchanges and, concomitantly, an asymmetrical diffusion of the particles \cite{Kaf00}.
This is achieved by allowing exchanges between particles on neighboring sites in the horizontal
direction to take place with the following rates:
\begin{eqnarray} \label{eq:ABC_rates}
AB \overset{q}{\underset{1}{\rightleftarrows}} BA~, \nonumber \\
BC \overset{q}{\underset{1}{\rightleftarrows}} CB~, \nonumber \\
CA \overset{q}{\underset{1}{\rightleftarrows}} AC~,
\end{eqnarray}
i.e. a $B$ particle and an $A$ particle to the right of the $B$ particle will swap places
with rate 1, whereas the same two particles will swap places with some rate $q<1$ in case their
order is different. Unbiased exchanges are of course recovered when $q=1$. The driven diffusive
motion that results when $q<1$ entails phase separation and the formation of ordered domains
in the form of vertical stripes \cite{Kaf00}
that are arranged in repetitions of the sequence $ABC$ where $A$ stands for a domain occupied
by a majority of $A$ particles.

For comparison, we also simulate the $ABC$ model on a ring composed of $L$ sites \cite{Eva98a,Eva98b}
where the same biased rates are used as for the horizontal direction in the two-dimensional model. 

Most of the data discussed below have been obtained for systems with $L=600$. We carefully checked that
this horizontal length is big enough so that it does not yield finite-size effects for the times
accessed in our simulations.

In all our simulations we consider fully occupied lattices where initially every species occupies
exactly one third of the lattice sites chosen at random. This is in fact a special situation
in one space dimension, as detailed balance is then fulfilled \cite{Eva98a}
so that the system relaxes to an equilibrium
steady state: in case of identical coverage of the lattice by each of the species,
a Hamiltonian with long-range interactions can be derived. However, as soon as the particle
densities are not the same for all three species, detailed balance is broken in one 
dimension and the steady state
is a non-equilibrium steady state. The phase transition between the ordered
phase for $q < 1$ and the disordered phase at $q=1$ (in the large volume limit) then becomes 
a non-equilibrium phase transition \cite{Cli03} which has been at the center of some of the recent studies
dealing with the $ABC$ model. For the two-dimensional model, however, detailed balance is
always broken, even when the three particle densities are identical,
and the steady state is always a non-equilibrium steady state \cite{Kaf00}. 

We are using a standard scheme for our simulations where we randomly select a neighboring pair of sites.
In case the particles occupying these sites belong to different species, they are exchanged with
the direction-dependent rates given above. We define a time step as $N$ such proposed updates,
where $N = L \times M$ is the total number of sites forming the lattice.

Every lattice site $\bf{x}$ can be characterized by a time-dependent three-state Potts variable 
$p({\bf x},t)$. Using that description, an exchange of particles occupying neighboring sites
corresponds to updating the values of the Potts variables characterizing these sites. In this
way it is also very easy to write down expressions for the quantities measured in our simulations.

A central quantity for our analysis is the equal-time spatial correlation
\begin{equation} \label{eq:spacecorr}
C({\bf r}, t) = \left< \frac{1}{N} \sum_{{\bf x}} \delta_{p({\bf x},t),p({\bf x}+{\bf r},t)}
\right> - \frac{1}{3}~,
\end{equation}
where $\delta$ is the Kronecker delta. The sum is over all lattice sites, and $\left< \cdots
\right>$ stands for an ensemble average over both initial configurations and noise realizations.
The subtraction by $\frac{1}{3}$ assures that in an infinite system $C({\bf r}, t)$ goes to zero 
when $\left| {\bf r} \right| \longrightarrow \infty$.

The space-time correlation (\ref{eq:spacecorr}) allows the extraction of a time-dependent
correlation length.
As we use different swapping rates in horizontal and vertical directions, the spatial
correlation, and therefore the correlation length, is direction dependent. 
For example, in the horizontal direction a time-dependent length $L_x(t)$
can be obtained in a standard way by determining the intersection of the normalized
correlation $C(\left| x \right|, t)/C(0, t)$ with a constant function $C_0$:
\begin{equation}
C(L_x(t), t)/C(0, t) = C_0~.
\end{equation}
We carefully checked that the qualitative features of $L_x(t)$ discussed below do not depend
on the chosen value of $C_0$. The data shown in the following have been obtained for $C_0=1/3$.
Alternatively, a time-dependent horizontal length can be obtained by measuring the average
horizontal extent of a connected cluster formed by particles of the same species. 
We carefully compared the two lengths and found
that both lengths are completely equivalent and show the same qualitative features.
A vertical length can be obtained from the vertical extent of these clusters.

Two-times quantities have been shown in many studies to be very valuable when investigating
the non-equilibrium properties of systems relaxing to a steady state \cite{Hen10}. The two-times
autocorrelation function
\begin{equation} \label{eq:autocorr}
C(t,s) = \left< \frac{1}{N} \sum_{{\bf x}} \delta_{p({\bf x},t),p({\bf x},s)}
\right> - \frac{1}{3}~
\end{equation}
compares the configurations at the waiting time $s$ with that at the observation time $t > s$.
Important insights can also often be obtained from two-times response functions. For the
model discussed here the only control parameter that can be used to perturb the
system is the swapping rate $q$. Therefore we use as our perturbation a sudden global change of
$q$ from some initial value $q_i$ to some final value $q_f$ at time $s$ after preparing
the system. For times $t > s$ we then monitor how the system relaxes to the final steady state
by measuring the time-dependent horizontal length $L_{x,p}(t,s)$ of the perturbed system. Of particular interest
is the difference \cite{Afz13}
\begin{equation} \label{eq:rep}
M(t,s) = \left| L_{x,p}(t,s) - L_x(t) \right|
\end{equation}
between the length measured in the perturbed system and that measured in the system where $q=q_f$
at all times.

All the quantities mentioned up to now do not provide direct
information on the behavior of individual particles. In order to probe the system at this
more microscopic level, we tag individual particles and follow their motion. From the resulting individual
trajectories we determine the displacement of each particle from its initial position as a function of time. In
the two-dimensional system particles of course move in both dimensions, but the vertical
motion is rather trivial as exchanges between neighboring particles of different types always take
place. For that reason we only discuss the mean displacement of the particles in the horizontal
direction:
\begin{equation}
d(t) = \left< \frac{1}{N} \sum\limits_{i=1}^N \left| x_i(t) - x_i(0) \right| \right>~,
\end{equation}
where $x_i(t)$ is the horizontal coordinate of the position vector ${\bf r}_i(t)$ of particle $i$
at time $t$, whereas $x_i(0)$ is the horizontal coordinate of the position of the 
same particle in the initial configuration.

\section{Time-dependent length scale}

As discussed in the previous section, in two space dimensions we can extract 
both a horizontal and a vertical length. 
Due to the anisotropy of the system, these lengths are  not identical. 
In order to directly compare with the
one-dimensional $ABC$ model, we mostly focus in the following on the length in the horizontal direction.

\begin{figure} [h]
\includegraphics[width=0.63\columnwidth]{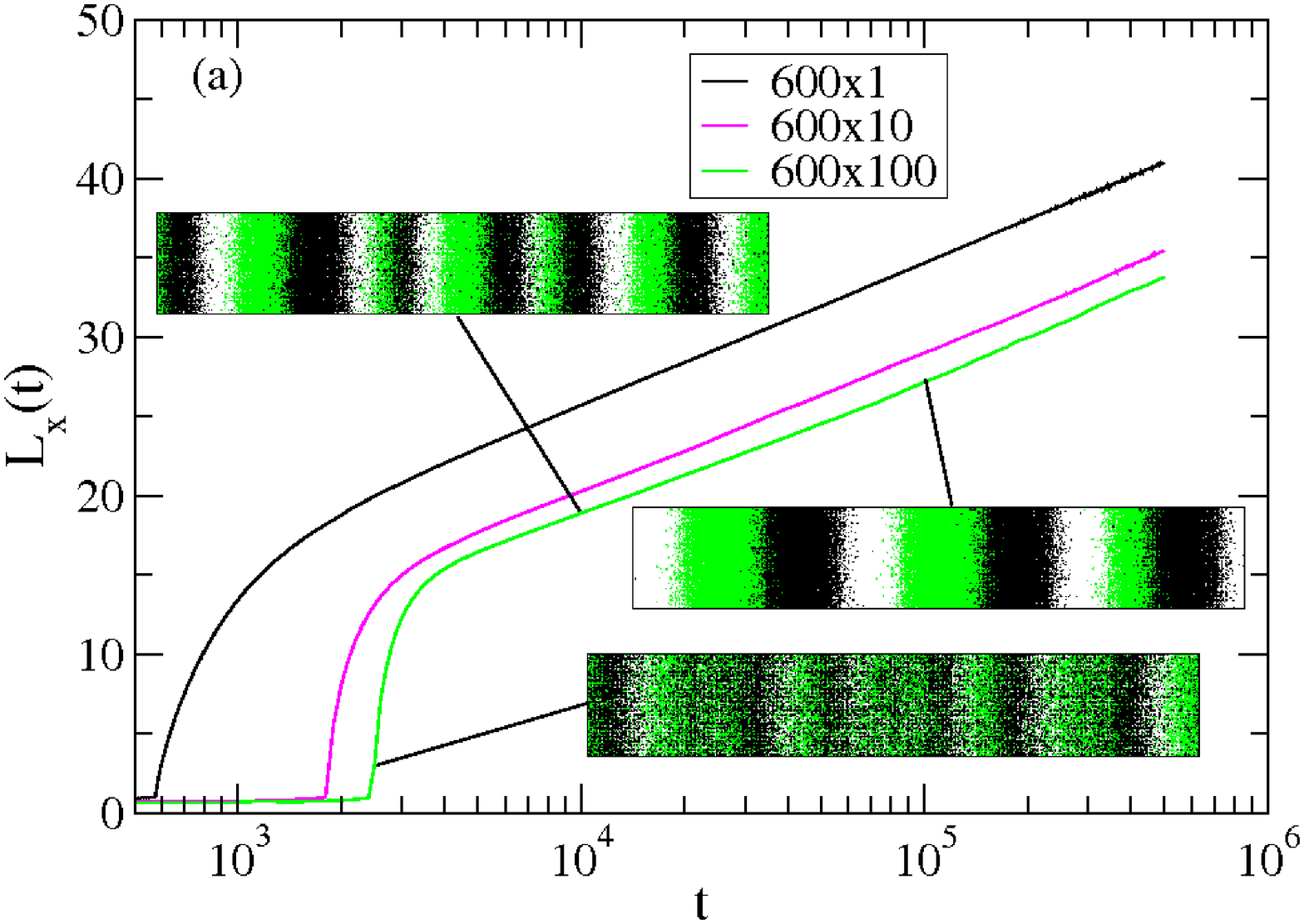}\\[0.3cm]
\includegraphics[width=0.64\columnwidth]{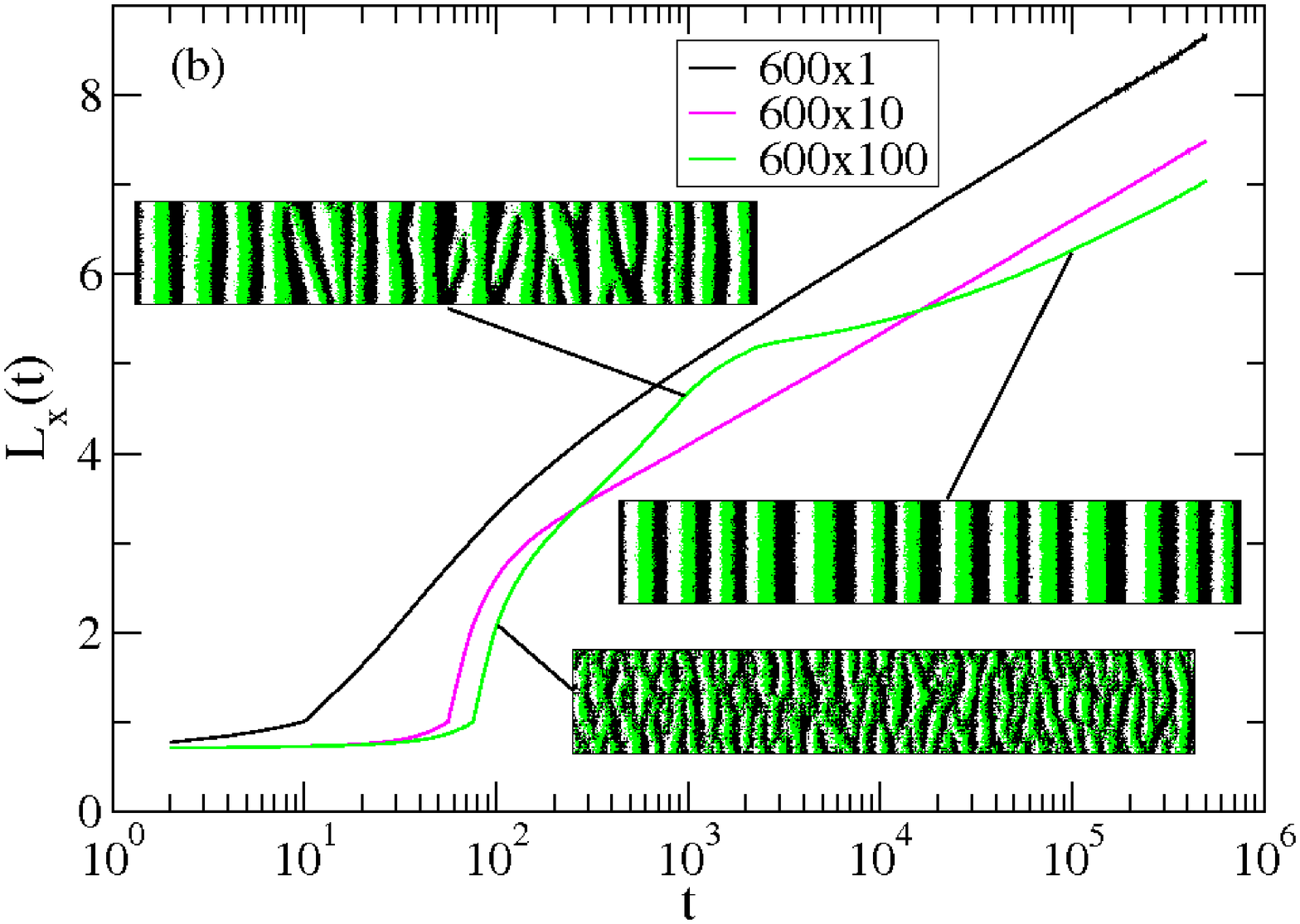}
\caption{\label{fig1} (Color online)
Time-dependent length $L_x(t)$ for (a) $q=0.8$ and (b) $q=0.2$. Systems with different extents in the 
$y$-direction are considered, ranging from 1 (top curve for large $t$) to 100 (bottom curve
for large $t$) lines. The length is obtained from the space-time correlation function
$C(\left| x \right|,t)$ in the horizontal direction. Some configurations are shown for the $600 \times 100$
system. For the larger systems the length is extracted after averaging over more than 1000 independent
realizations, whereas for the one-dimensional system the ensemble average is performed over 
16000 independent runs.
}
\end{figure}

\begin{figure} [h]
\includegraphics[width=0.85\columnwidth]{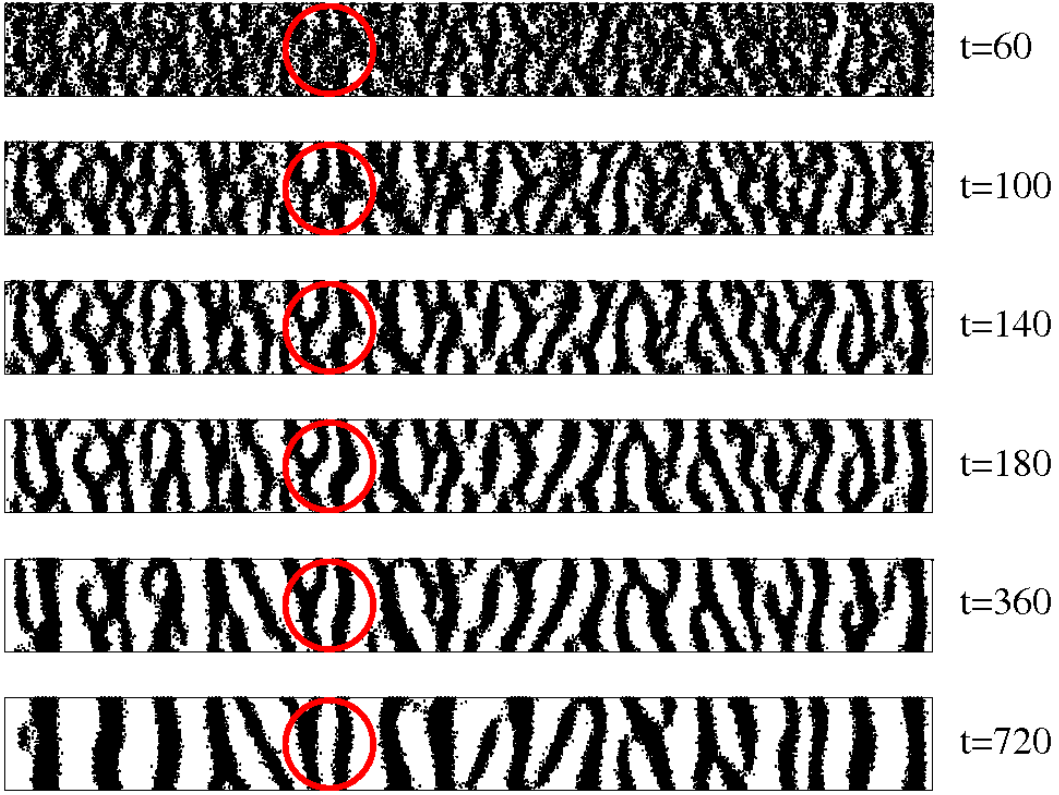}
\caption{\label{fig2} (Color online)
Snapshots of a system with $q=0.2$ composed of $600 \times 100$ sites for six different times 
between $t=60$ and $t=720$. Only one of the three particle
types is shown for clarity. The times shown span the initial
regime of domain formation, the regime where the domains merge and expand in vertical direction, and the
regime where defects are eliminated and the straight stripes are forming, 
after which the system enters the logarithmic
coarsening regime. The circle highlights one instance where the misalignment of merging domains and
the following expansion in the 
vertical direction yields bended domains as well as a $Y$-shaped connection.
}
\end{figure}

Fig.~\ref{fig1} shows the horizontal length $L_x(t)$ for systems of up to 100 lines and two very different values of $q$,
namely $q=0.8$ in Fig.\ \ref{fig1}a and $q=0.2$ in Fig.\ \ref{fig1}b. The value
$q=0.8$ is close to the value $q=1$ for unbiased exchanges. This closeness to $q=1$
gives raise to a very slow initial onset of the ordering process, and it takes for a system composed
of 10 or more lines around 2000 time steps before a growth in $L_x(t)$ can be observed. This is different
for the case of strong bias, $q=0.2$, where ordering emerges already after less than 100 time steps.
Once ordering sets in for $q=0.8$, it rapidly yields the formation of ordered stripes (see
the configurations in Fig.\ \ref{fig1}a for the system with 100 lines). These stripes initially contain
a large number of particles from the minority species. When the domains are well formed, a coarsening
process sets in that reveals itself by a logarithmic growth of $L_x(t)$. For systems with $q=0.2$ 
and only a few lines, this behavior is qualitatively very similar to the behavior encountered for $q=0.8$,
the main differences being the earlier onset of ordering and the rapid formation of a large number of
well ordered domains in form of stripes. Once these domains are formed, they coarsen very slowly and grow only
logarithmically with time. Interestingly, for systems with a larger vertical extent (see the data for the
$600 \times 100$ system in Fig.\ \ref{fig1}b) an intermediate regime shows up after the onset of ordering
but before the domain coarsening. The nature of this regime, which is characterized by an accelerated
increase of $L_x(t)$ followed by a strong slowing down of the growth process,
is revealed when inspecting the typical configurations
shown in  Fig.\ \ref{fig1}b and Fig.\ \ref{fig2}. In the early stages of the ordering process, clusters
merge and grow in vertical direction until they reach from bottom to top. This vertical growth, which
gives raise to a transient algebraic growth regime, yields initially some defects
that need to be eliminated before entering the logarithmic coarsening regime. Consequently, the
asymptotic growth regime sets in much later than for the systems with only a few lines.

In order to gain a more microscopic picture of the processes taking place in the pre-asymptotic regimes,  
we show in Fig. \ref{fig2} six different snapshots, with times between $t=60$ and $t=720$, for a
system with $600 \times 100$ sites and $q=0.2$. Only one of the three particle types
is shown, as this allows to better see the processes taking place in
the system. The initial ordered clusters develop
at random positions and rapidly merge into anisotropically ordered domains where the growth in vertical direction
is much faster than in horizontal direction. When merging, these vertically rapidly growing domains 
are not always well aligned,
which yields defects as for example bended stripes or $Y$-shaped connections. The red circles in Fig. \ref{fig2}
indicate the formation and subsequent evolution of some defects. Once these imperfect stripes
are formed, a regime sets in where the defects are eliminated. At $t=720$ only few defects are
left and most of the stripes, which have become much straighter, extend through the whole system in
vertical direction. The asymptotic logarithmic growth regime is fully accessed only after all 
these defects have been eliminated.

\begin{figure} [h]
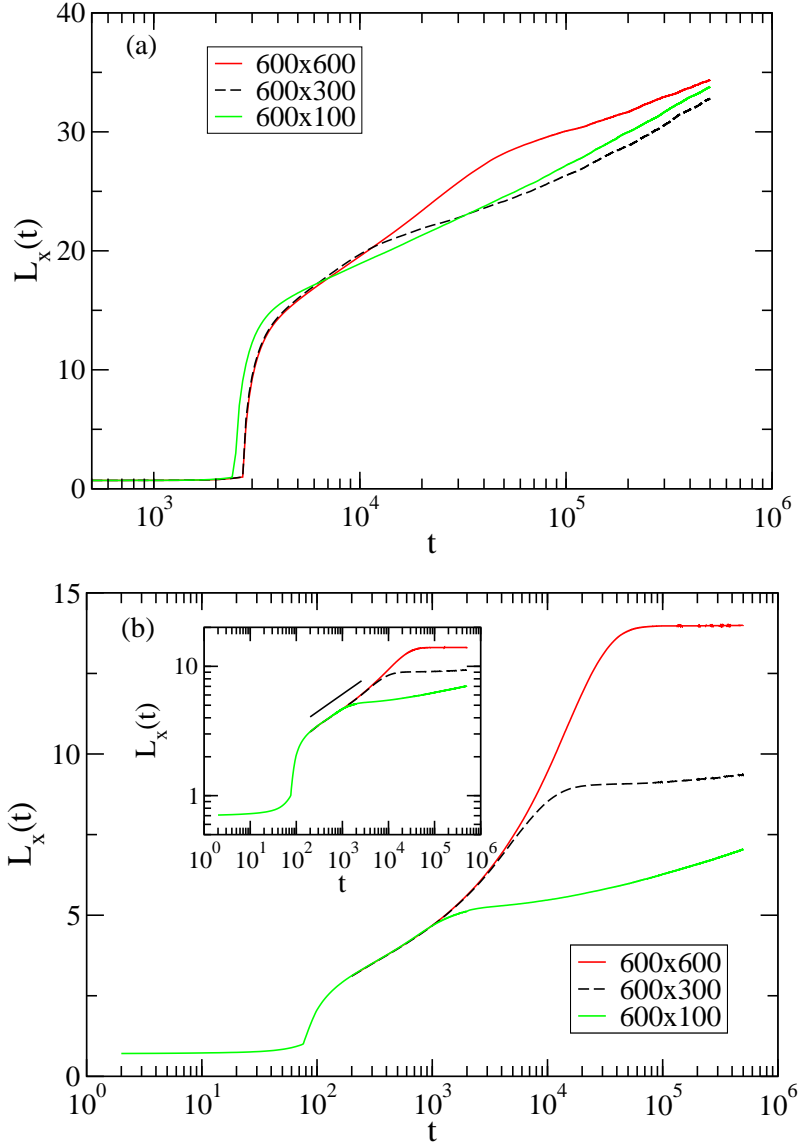

\includegraphics[width=0.63\columnwidth]{figure3a.eps}\\[0.3cm]
\includegraphics[width=0.64\columnwidth]{figure3b.eps}
\caption{\label{fig3} (Color online)
The same as in Fig. \ref{fig1}, but now for systems with a larger extent in $y$-direction.
The data result from averaging over at least 1000 independent runs. The inset in (b) shows
the same data as in the main panel, but now in a log-log plot. The full solid line indicates 
the intermediate algebraic growth regime with an exponent 0.25(1).
}
\end{figure}

Fig. \ref{fig3} shows the behavior of $L_x(t)$ when further increasing the number of lines. We first note
from Fig. \ref{fig3}b that for $q=0.2$ the intermediate regime lasts longer the larger the 
vertical extent of the system is, as it takes longer to form the parallel stripes
spanning the system. The approximate power-law growth connected to this regime is revealed in the log-log plot shown
in the inset. The full solid line indicates that this growth is governed by an effective exponent
of 0.25(1). For $q=0.8$, see Fig.~\ref{fig3}a, the intermediate regime also emerges for larger systems in $y$-direction.
In fact the same regimes exist for all values of $q$ and only the characteristic
extent in $y$-direction
needed for the existence of the intermediate regime depends on $q$.

\begin{figure} [h]
\includegraphics[width=0.63\columnwidth]{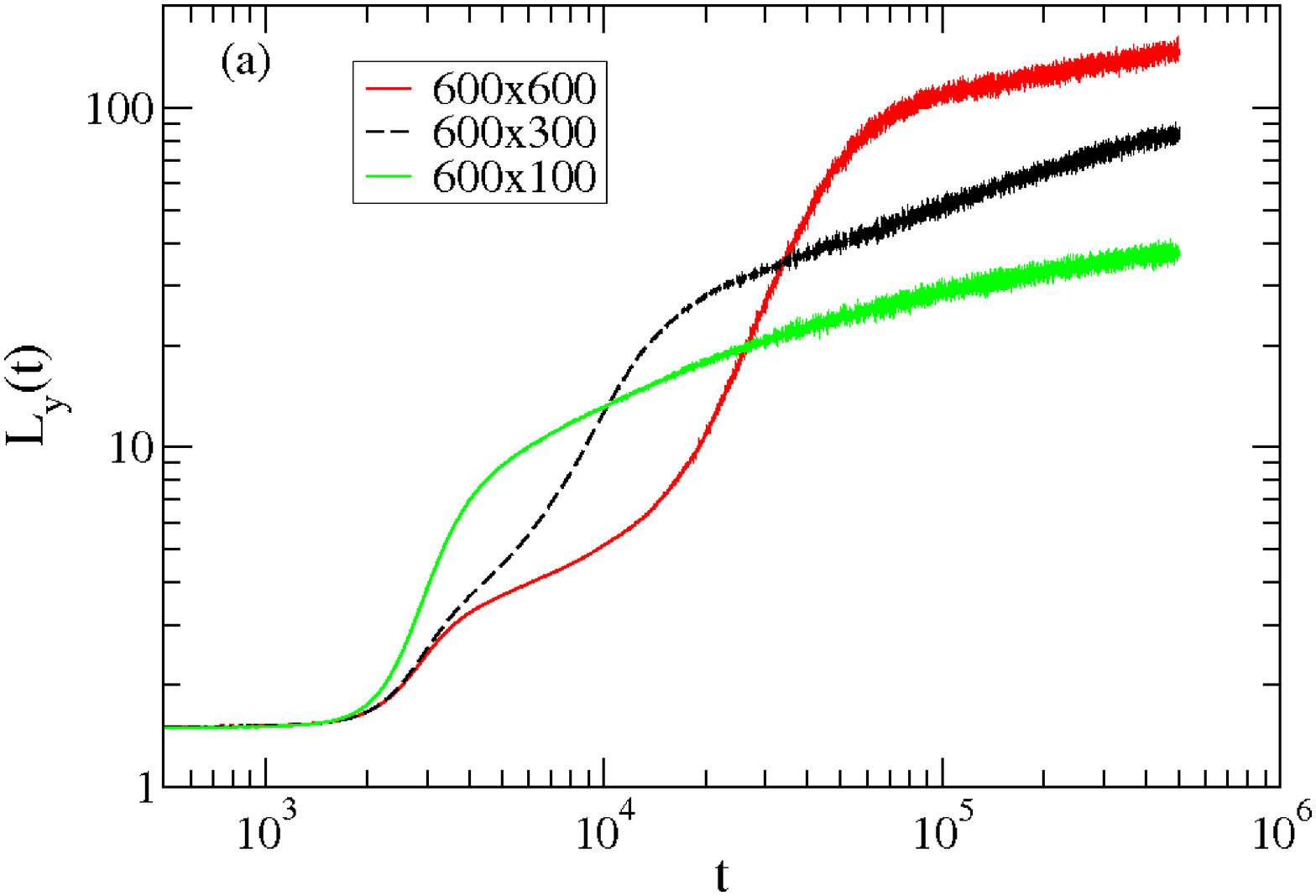}\\[0.3cm]
\includegraphics[width=0.64\columnwidth]{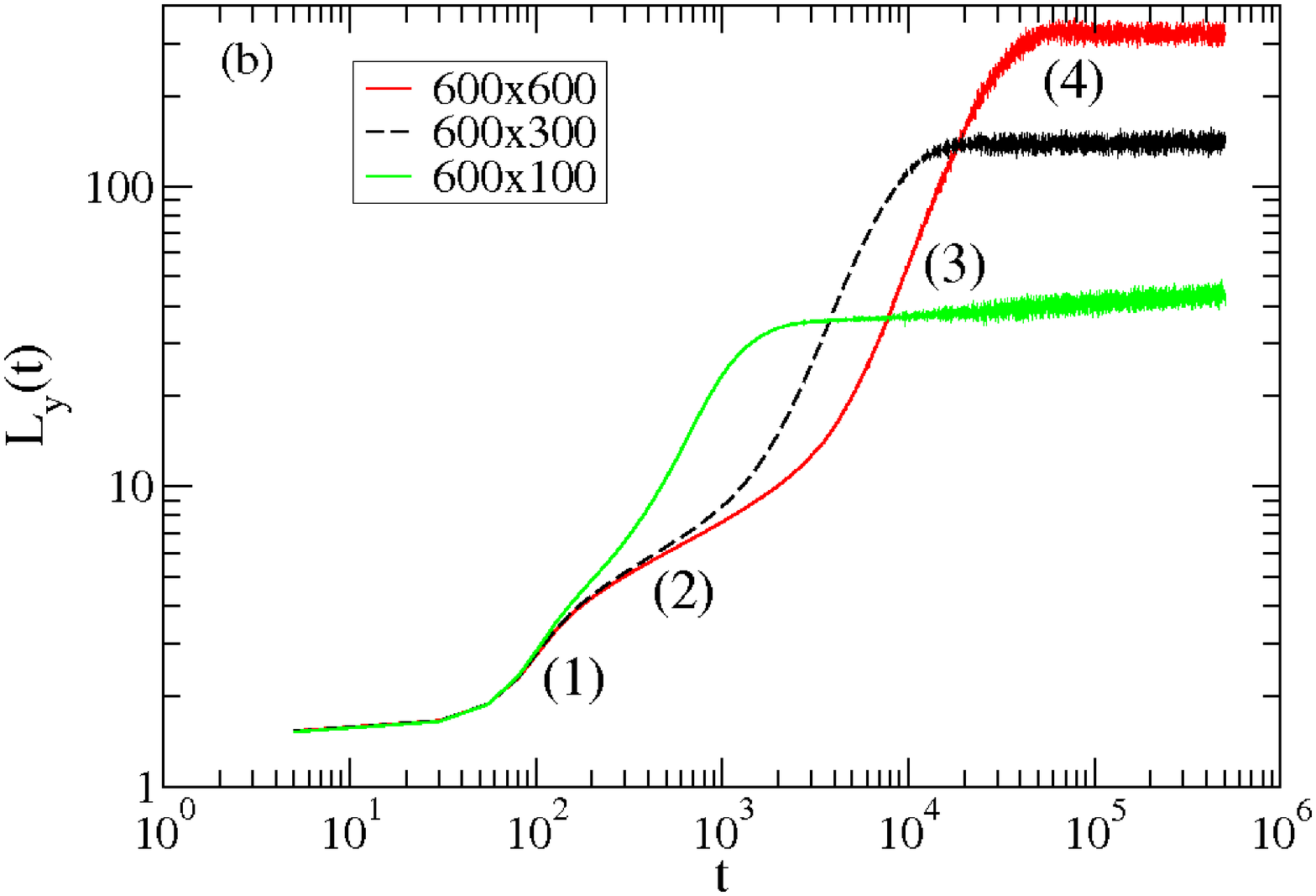}
\caption{\label{fig4} (Color online)
Time evolution of the vertical length $L_y(t)$ for (a) $q=0.8$ and (b) $q=0.2$. The systems all have 600
sites in horizontal direction, but different extents in $y$-direction. The different regimes indicated by
numbers in (b) for the system with $600 \times 600$ spins are discussed in the main text.
}
\end{figure}

The different regimes identified from our analysis of $L_x(t)$ also reveal themselves through
characteristic features in the vertical length, as shown in Fig. \ref{fig4} where we plot
the average vertical extent of connected clusters. 
Focusing first on Fig. \ref{fig4}b, we note that 
the initial domains grow rapidly in the early time regime (1),
before entering the intermediate regime (2) where the merging of ordered clusters yields
defect structures that needs to be eliminated. The growth regime (3)
then corresponds to the elimination of these defects and the straightening of the stripes. 
The asymptotic regime (4), characterized
by a logarithmic growth in horizontal direction, reveals itself in the following very slow
increase of $L_y(t)$. All these features are also observed in Fig. \ref{fig4}a for $q=0.8$, albeit
less pronounced and separated by more gradual transitions.

In \cite{Eva98b} and \cite{Kaf00} interface models with simplified dynamics were proposed 
both in one and two space dimensions in order 
to capture the long time behavior. These models assume instantaneous moves of particles
from one domain to another of the same type, with a rate that depends on $q$ as well as
on the width of other stripes located between these two domains.
In doing so, one neglects that in the original $ABC$ model particles need a finite amount of time to cross these intermediate stripes.

\begin{figure} [h]
\includegraphics[width=0.85\columnwidth]{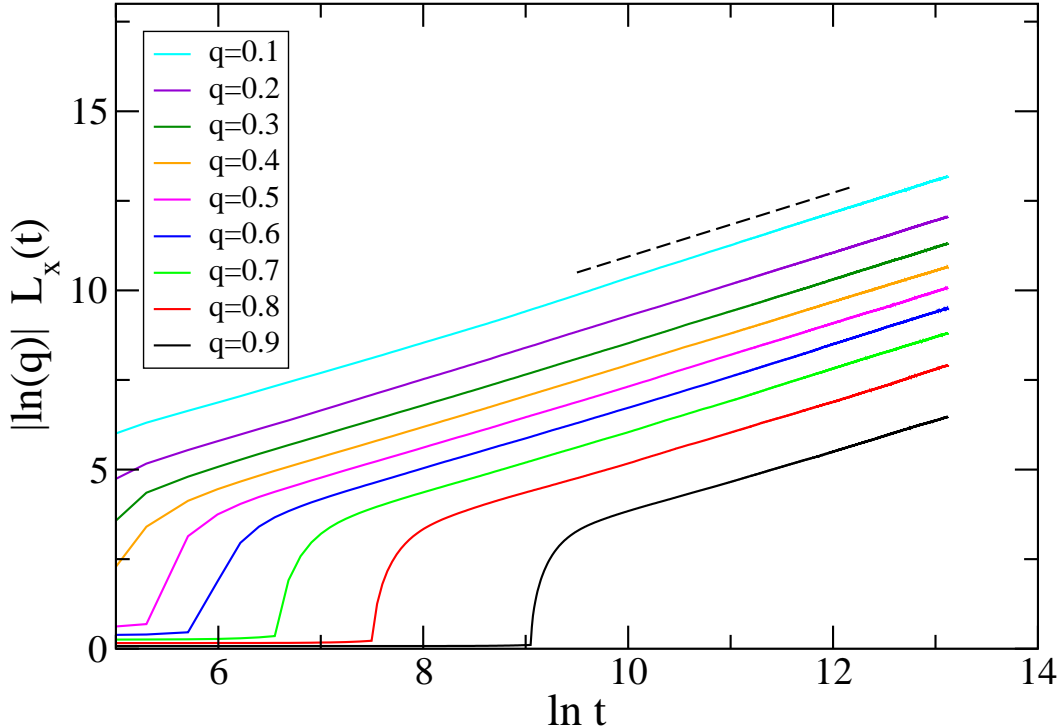}
\caption{\label{fig5} (Color online)
Time-dependent length $L_x(t)$ vs $\ln t$ for various values of the bias $q$ in a system composed
of $600 \times 10$ sites. The values of $q$ increase from top to bottom. 
Plotting $|\ln(q)| L_x(t)$ on the $y$-axis yields in all cases curves with
slopes $\kappa = 0.89(2)$.
}
\end{figure}

For the $600 \times 1$ and $600 \times 10$ systems we are able to access in the microscopic model the asymptotic growth regime for all
values of $q$ between 0.9 and 0.1. In agreement with what has been observed for the interface model \cite{Eva98b,Afz13}, we find
also for the microscopic model that the plot of $L_x(t)$ versus $\ln t$ yields a slop that depends on $q$ in the
following way:
\begin{equation}
L_x(t) = \kappa \,  \ln t / \left| \ln q \right|
\end{equation} 
with $\kappa = 0.89(2)$ for both system sizes. 
This is illustrated in Fig. \ref{fig5} for the system containing $600 \times 10$ sites.
The value of $\kappa$ is much smaller than the value $\kappa \approx 2.0$ found for the interface
model in one dimension \cite{Afz13}, in agreement with the expectation that the simplified dynamics in that model leads to a change in the
value of the unit of time.

\begin{figure} [h]
\includegraphics[width=0.85\columnwidth]{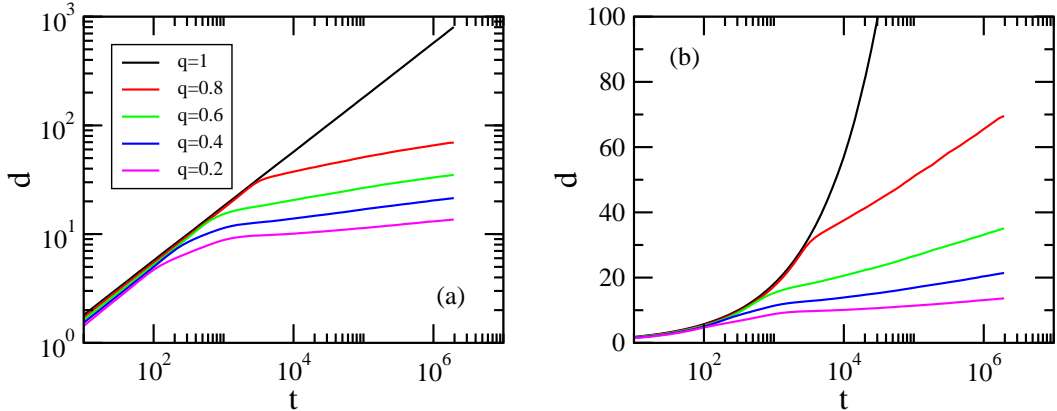}
\caption{\label{fig6} (Color online)
Mean displacement in horizontal direction
of particles in a system composed of $9000 \times 100$ sites
in (a) a log-log plot and (b) a log-linear plot. 
The values of $q$ decrease from top to bottom.
Results for different values of $q$ are shown. For $q=1$
a square-root increase of $d$ is observed, 
as expected for an unbiased random walk. For $q < 1$ a crossover to a regime where the 
displacement only grows logarithmically with time is encountered. These data result from averaging over 16 independent
runs.
}
\end{figure}

Taking a more microscopic point of view, we can record the trajectories of the individual particles
and determine in this way their typical motion. Fig. \ref{fig6} shows the mean displacement
in horizontal direction of particles moving in systems with 100 lines. As is the case for $L_x(t)$, 
two well-separated regimes can also be identified for large $q$ values when studying this quantity,
whereas for smaller $q$ values indications of an additional intermediate regime are seen. Initially, the
mean displacement for every $q$ displays the same square-root behavior as for $q=1$. 
This indicates an unbiased diffusion regime at short times before the formation of stripes sets in. For large values
of $q$, a crossover to a logarithmic time dependence takes place once the stripes are formed. Note that this
transition is very sharp for $q=0.8$. For the smaller values of $q$, see for example the data for $q=0.2$, a more
gradual transition is observed, in agreement with the more complicated behavior of $L_x(t)$ for that case.
This gradual transition is related to the elimination of defects in the horizontal direction and the
subsequent straightening of the stripes. During this process particles can move more freely in vertical
direction than in horizontal direction, as due to the defect structures
not every particle in a stripe is surrounded in vertical direction by particles of the same type.
Consequently, it happens that a particle at the borderline region of 
an imperfect stripe lands after moving only a few steps in vertical
direction in another layer where it finds itself the midst of a domain 
of another species, thus allowing it to then move in the horizontal
direction. As shown by our data, this process yields a displacement in
horizontal direction that is slower than diffusive and faster than
logarithmic.

\begin{figure} [h]
\includegraphics[width=0.85\columnwidth]{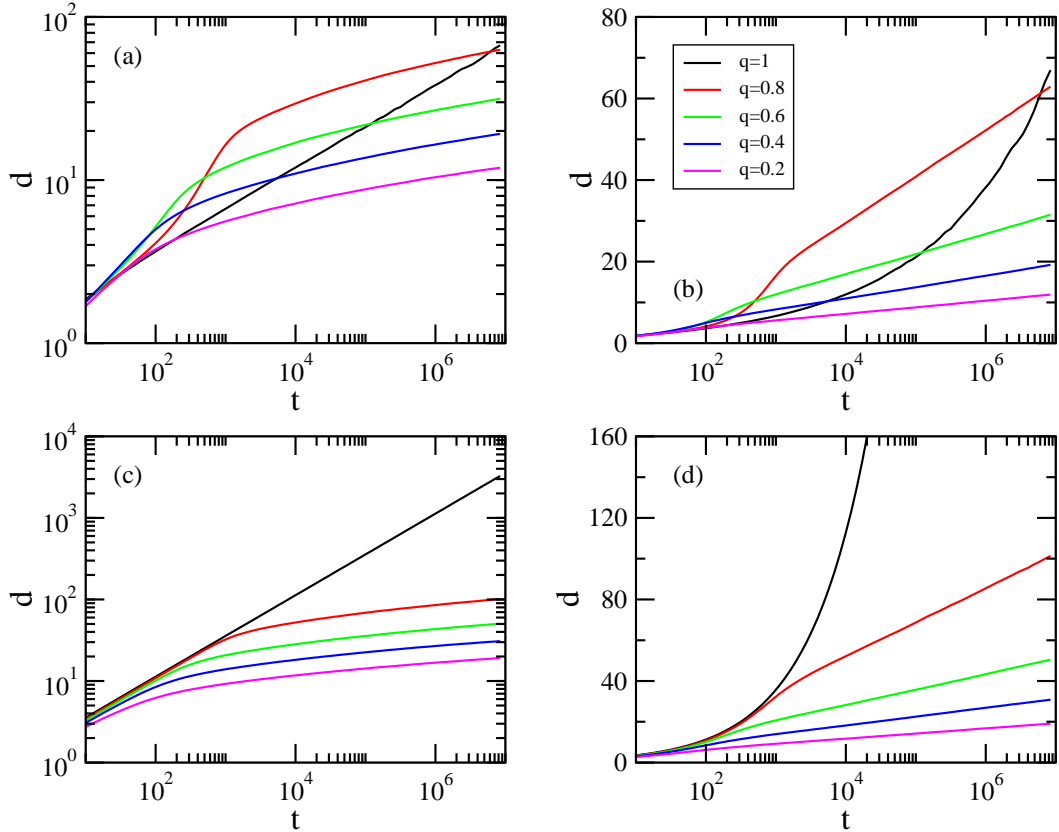}
\caption{\label{fig7} (Color online)
Mean displacement of particles on the one-dimensional ring with 9000 sites
in (a,c) a log-log plot and (b,d) a log-linear plot. 
At very large times, the values of $q$ decrease from top to bottom.
In (a,b), following the rules of our model,
a particle is not exchanged with a neighboring particle of the same type, yielding for unbiased exchanges
with $q=1$ an effective sub-diffusive behavior $d \sim t^{1/4}$. When also allowing for exchanges between
particles of the same type, see (c,d), the square-root increase of a random walk prevails. Irrespective whether or not
exchanges between like particles are allowed, a crossover to a logarithmic increase is observed for $q < 1$.
The curves result from an average over 200 independent runs.
}
\end{figure}

The data shown in Fig. \ref{fig7}a and \ref{fig7}b for the ring reveal a behavior that at first
look seems odd. Indeed for $q=1$ one does not observe the free diffusion of particle, but instead the 
particles move sub-diffusively, with a mean displacement $d \sim t^{1/4}$. In order to understand this,
let us recall that in the definition of the ABC model, see Section II, exchanges between particles of the
same type are not assumed. In fact, for all aspects discussed in previous papers as well as for all the time-dependent
quantities studied in our work, with the exception of the mean displacement, it is completely irrelevant
whether we allow exchanges of same particles or not. However, for individual trajectories this does matter, as without
these exchanges particles get stuck behind others of the same type and can only move if a particle of a different 
type comes by and replaces the particle on the neighboring site. Indeed, when allowing for these additional exchanges, we do recover
free diffusion for $q=1$ also in one dimension, and for $q < 1$ a crossover between that early time behavior and
a logarithmic regime is observed, see Fig. \ref{fig7}c and \ref{fig7}d. 

In the two-dimensional system shown in Fig. \ref{fig6} particles get rarely stuck as they can diffuse both in $x$- and
$y$-direction. Consequently, we have for $q=1$ the expected square-root behavior of a random walk even in absence of exchanges
between particles of the same species.

\begin{figure} [h]
\includegraphics[width=0.63\columnwidth]{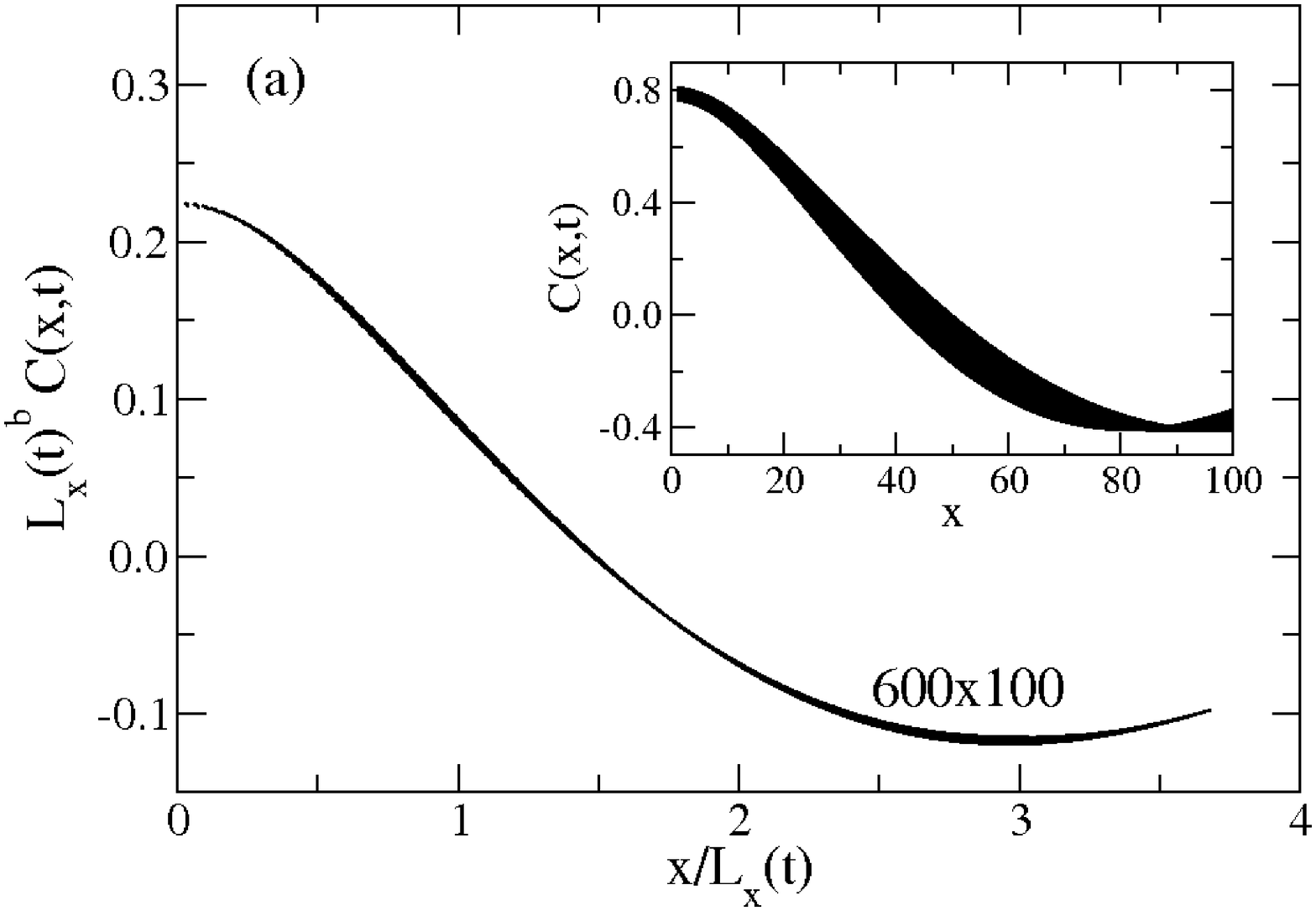}\\[0.3cm]
\includegraphics[width=0.63\columnwidth]{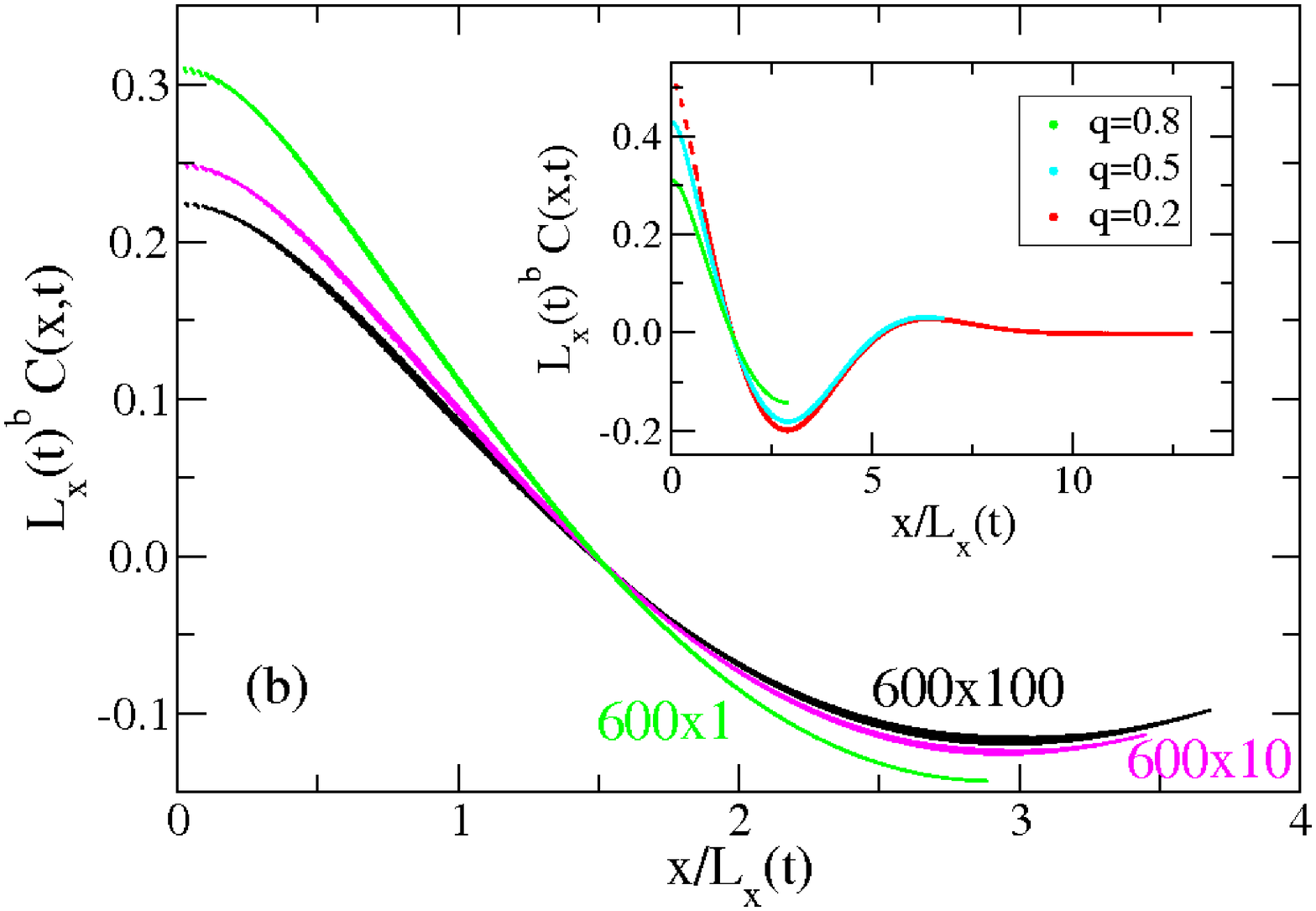}
\caption{\label{fig8} (Color online)
Scaling of the equal-time correlation function for $q=0.8$ and different system sizes. Data ranging from
$t=100000$ to $t=500000$ are included. (a) Rescaled data for a system with $600 \times 100$ sites.
The inset shows the unscaled data. (b) Comparison of the scaling functions for three different
system sizes. The inset shows the scaling of the
correlator for a ring with 600 sites and three different values of the swapping rate $q$. The values of the
scaling exponent $b$ that achieve the best data collapse are gathered in Table \ref{tab1}. 
For the larger systems these data result from averaging over more than 1000 independent
realizations, whereas for the one-dimensional system the ensemble average is performed over 
16000 independent runs.
}
\end{figure}

For systems governed by a time-dependent length scale $L_x(t)$ one expects from general considerations \cite{Bra93} that
the single-time correlator $C({\bf r},t)$ exhibits 
a simple scaling behavior of the form \cite{Hen10}
\begin{equation} \label{eq:C_t_scal}
C({\bf r},t) = \left( L_x(t) \right)^{-b} f\left( \frac{\left| {\bf r} \right| }{L_x(t)} \right)
\end{equation}
for sufficiently large times $t \gg t_{micro}$, where $t_{micro}$ is a microscopic reference time such that 
$L_x(t_{micro})$ is larger than any microscopic length scale. In the present case $t_{micro}$ has to be larger
than the typical time needed for the formation of ordered stripes. As shown in Fig. \ref{fig8}a, once the logarithmic
growth regime has been fully accessed, dynamical scaling prevails and a data collapse is achieved when assuming
the scaling (\ref{eq:C_t_scal}). The values of the scaling exponent $b$ are gathered in Table \ref{tab1} for
three different vertical extents and three different values of the swapping rate $q$. Inspection of the table reveals a weak
dependence on the number of lines in the system. In addition, the value of $b$ changes with $q$. A similar dependence is also seen
for the scaling function itself, see the inset in Fig. \ref{fig8}b. 
It follows that for the $ABC$ model scaling functions
and scaling exponents depend on system parameters.
This is similar to what is seen in other systems with
logarithmic growth \cite{Par10,Cor11,Cor12,Par12,Cor13}.

Finally, we note from Fig. \ref{fig8}b that the scaling functions for different extents in
vertical direction are not identical for the finite times accessed in our
simulations. This can be viewed as a 'finite-size' effect, and the 
dependence of the scaling function (and, of course, of $L_x(t)$) on the vertical direction
will vanish in the large volume limit and long time limit.

\begin{table}
\begin{tabular}{|c|c|c|c|} \hline
 & $600 \times 1$ & $600 \times 10$ &  $600 \times 100$ \\ \hline
$q=0.2$ & $-0.07(1)$ & $-0.12(1)$ & $-0.12(2)$ \\
$q=0.5$ & $-0.11(1)$ & $-0.15(1)$ & $-0.21(2)$ \\
$q=0.8$ & $-0.16(1)$ & $-0.22(1)$ & $-0.25(2)$ \\ \hline
\end{tabular}
\caption{Values of the scaling exponent $b$ for three different systems and three different values of the swapping rate $q$.
\label{tab1}}
\end{table}

\section{Two-times quantities}

Two-times quantities are used in many studies in order to elucidate relaxation processes and aging
phenomena far from stationarity \cite{Hen10}. As the discussion of the equal-time correlation function has
established the presence of dynamical scaling in our system, we expect the following scaling \cite{Hen10} for the two-times
autocorrelation function $C(t,s)$, see Eq. (\ref{eq:autocorr}):
\begin{equation} \label{eq:C_t_s_scal}
C(t,s) = \left( L_x(s) \right)^{-b} F\left( \frac{L_x(t)}{L_x(s)} \right)
\end{equation}
where the exponent $b$ should be the same as for the equal-time correlator (\ref{eq:C_t_scal}). This scaling form assumes
that $L_x(t)$, $L_x(s) \gg L_x(t_{micro})$ as well as $L_x(t)-L_x(s) \gg L_x(t_{micro})$. This last condition can be difficult
to fulfill for a system with logarithmic growth, and sizeable finite-time corrections could spoil the
expected scaling. 

\begin{figure} [h]
\includegraphics[width=0.85\columnwidth]{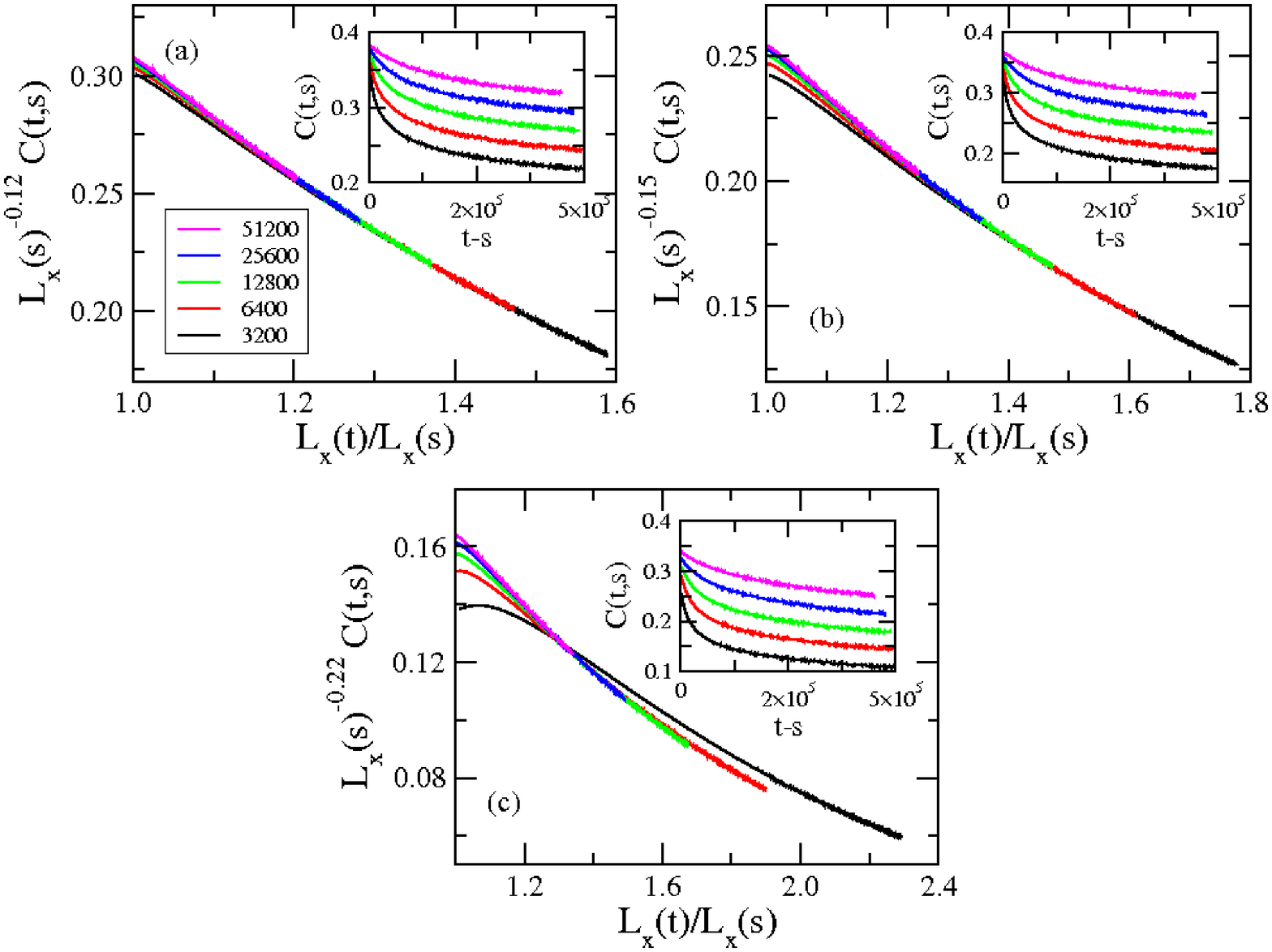}
\caption{\label{fig9} (Color online)
Scaling of the two-times autocorrelation function $C(t,s)$ for a system composed of $600 \times 10$
spins, with (a) $q=0.2$, (b) $q=0.5$, and (c) $q=0.8$. The insets show the autocorrelation as function
of the time difference $t-s$ and reveal the aging properties of that quantity. The waiting times
decrease from top to bottom. In all cases a good data
collapse is achieved for the larger waiting times $s$ when assuming the scaling form (\ref{eq:C_t_s_scal}),
with $b$ given in Table \ref{tab1}. These data result from averaging over 8000 runs.
}
\end{figure}

Fig. \ref{fig9} explores the scaling (\ref{eq:C_t_s_scal}) for a system with $600 \times 10$ sites and three
different swapping rates $q$. These three panels are also representatives for other system sizes and other
values of $q$. Besides large waiting times $s$ for which we clearly are in the logarithmic growth
regime, we also use rather short ones where in some cases the intermediate regime has not yet been left.
We first remark that in all studied cases dynamical scaling (\ref{eq:C_t_s_scal}) indeed prevails for the larger waiting times
and $L_x(t)/L_x(s)$ not too close to 1. This is indeed achieved with the exponents listed in Table \ref{tab1}.
On the other hand, finite-time corrections, which become less important for increasing $s$,
are readily identified in the vicinity of $L_x(t)/L_x(s) \approx 1$. For the case
$q=0.8$ shown in panel (c) the shortest waiting time $s=3600$ is obviously not yet in the scaling regime. Doing
the same analysis for larger systems, say ones composed of $600 \times 100$ sites, we have a data collapse only 
for waiting times for which the system left the crossover regime and entered the asymptotic
logarithmic regime with ordered stripes.

\begin{figure} [h]
\includegraphics[width=0.85\columnwidth]{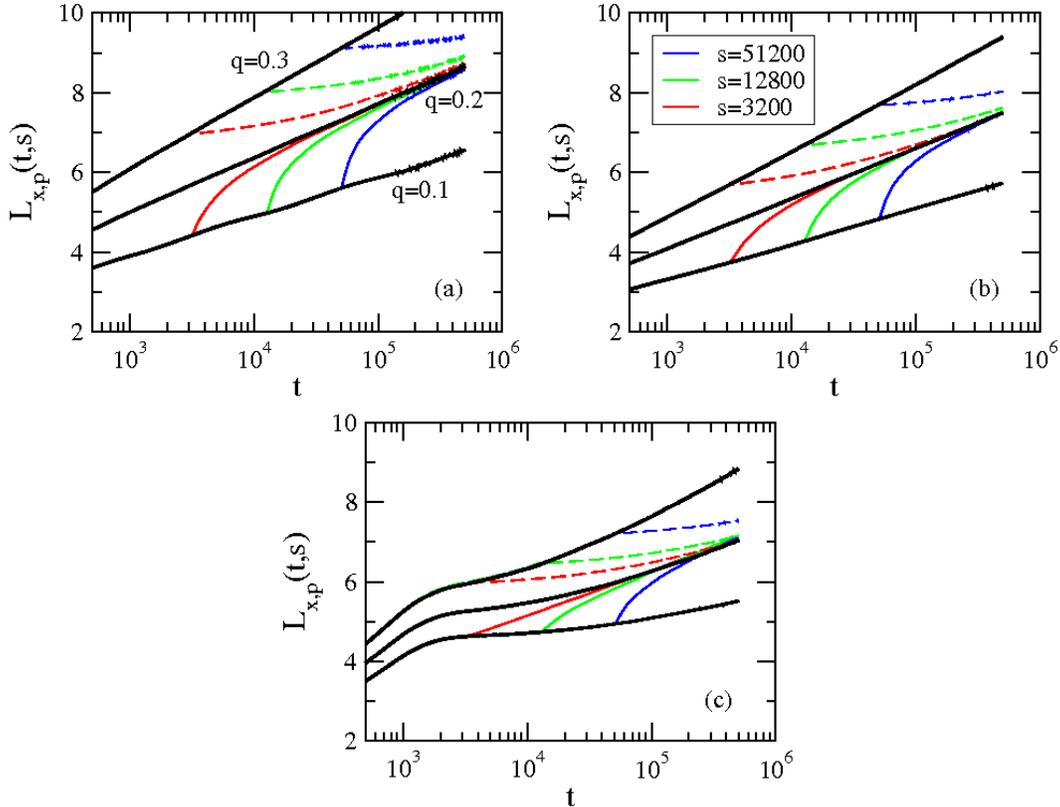}
\caption{\label{fig10} (Color online)
Time-dependent length when changing at the waiting time $s$ the swapping rate from $q_i=0.3$ to $q_f=0.2$ (dashed colored
lines) and from $q_i=0.1$ to $q_f=0.2$ (full colored lines). 
The waiting times increase from left to right. The system sizes are (a) $600 \times 1$, (b)
$600 \times 10$ and (c) $600 \times 100$. The full black lines indicate the growing length for a system
where $q$ is fixed at 0.3, 0.2, or 0.1 (from top to bottom). These data result from averaging over at least
1600 independent runs.
}
\end{figure}

Whereas the two-times correlation function contains information on the time evolution of a system by
comparing configurations at different times since its initial preparation, a two-times
response function provides insights in how a system that still evolves reacts to a perturbation.
A variety of protocols can be designed in order to probe the response of a system to a change of system
parameters. As already discussed in Section II, we consider in the following sudden changes in the
biased exchange rate $q$ and monitor how the perturbed system relaxes to the same state as that
reached by a control system that evolves all the time at that final value of $q$.

Fig. \ref{fig10} compares the length after the perturbation, $L_{x,p}(t,s)$, with that of the unperturbed
system for two cases: (1) $q_i =0.3$ and $q_f =0.2$ (dashed lines) and (2) $q_i=0.1$ and $q_f = 0.2$ (full lines).
For systems with $600 \times 1$ and $600 \times 10$ sites the perturbations happen after the
formation of the stripes. On the other hand, for the system with $600 \times 100$ sites shown in panel (c) the perturbations
take place close to the end of the intermediate regime but before the logarithmic growth is fully established. 
Qualitatively the responses
for all three system sizes are very similar, though. In all cases the perturbed length tends to the corresponding length of
the control system. This approach is rather fast 
for an increase of $q$ (in the present case from 0.1 to 0.2), but takes much longer for a decrease of $q$.
When increasing $q$ the domains just after the change are too compact when compared to the domains
that grow at the fixed value $q=0.2$.
On the other hand, when considering the case where $q$ is decreased, the
domains after the change are less dense than what should be the case for a $q=0.2$ system.
Although $L_x(t=s)$ is larger for $q=0.3$ than that for
$q=0.2$, $L_x(t)$ does not decrease after the quench, see the dashed lines
in Fig. \ref{fig10} that remain almost flat. Once the domains are
formed, their size can not be reduced in a sizeable manner; the domain growth
will be slowed down until the time $t_0$ when $L_x(t=t_0)$ for the system evolving at the fixed rate
$q=0.2$ is comparable to $L_x(t=s)$ for the $q=0.3$ system.

\begin{figure} [h]
\includegraphics[width=0.85\columnwidth]{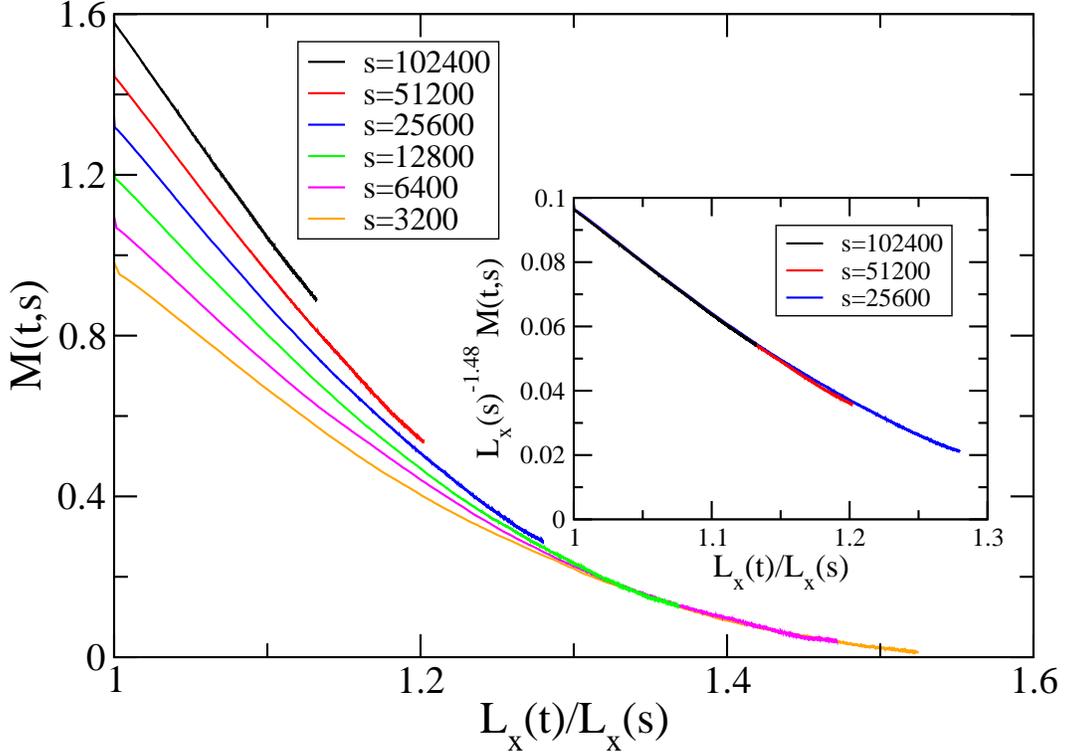}
\caption{\label{fig11} (Color online)
Response function $M(t,s)$ for a system composed of $600 \times 10$ particles 
after the diffusion bias has been changed from $q_i =0.3$ to $q_f = 0.2$ at 
time $s$. The waiting times $s$ decrease from top to bottom. The inset shows a reasonable scaling 
$M(t,s) = \left( L_x(s) \right)^{-a} F_M\left( \frac{L_x(t)}{L_x(s)} \right)$
for the three largest waiting times with a scaling exponent $a=-1.48$.
}
\end{figure}

In principle, the two-times response of a system to an infinitesimal instantaneous perturbation
should display a scaling behavior similar to that of the two-times
correlation function, see Eq. (\ref{eq:C_t_s_scal}). 
In our case, however, the perturbation, i.e. the change in $q$, is a large one, with $\left| q_f - q_i \right| = 0.1$.
One should also note that a scaling behavior can not be expected when the perturbation
takes place in a regime where $L_x(t)$ displays a crossover between different types of behaviors. Inspection
of Fig. \ref{fig10}c, for example, reveals for the change from $q_i=0.1$ to $q_f = 0.2$ a rather different behavior of
$L_{x,p}(t,s)$ for $s=3600$ than for $s=51200$ (see full lines), and dynamical scaling
can not be realized.

For an increase of $q$ we were unable to achieve a data collapse using as scaling variable
$L_x(t)/L_x(s)$. This is different from when $q$ is decreased, however, as dynamical scaling can
be identified in this case. Fig.~\ref{fig11} discusses as an example the change of $q$ from 0.3 to 0.2 for a system composed
of $600 \times 10$ sites. The difference $M(t,s)$ between the perturbed length $L_{x,p}(t,s)$ and the length $L_x(t)$ of the
control system evolving with $q=0.2$, see Eq. (\ref{eq:rep}), provides a quantity that displays aging as it does
not merely depend on the time difference $t-s$. $M(t,s)$ goes to zero for long times when the ordered stripes 
of the perturbed system get similar to those of the unperturbed system. Inspection of Fig. \ref{fig11} 
reveals the emergence of two different regimes: a regime where $M(t,s)$ is small (i.e. the perturbed
length is close to the unperturbed one) and only depends on $L_x(t)/L_x(s)$, and a regime 
close to $L_x(t)/L_x(s) \approx 1$ where $M(t,s)$ is large and shifts
to higher values for increasing waiting times $s$. 
In that regime a good scaling can be achieved when assuming that
\begin{equation}
M(t,s) = \left( L_x(s) \right)^{-a} F_M \left( \frac{L_x(t)}{L_x(s)} \right)~,
\end{equation}
see the inset in Fig. \ref{fig11}. We remark that immediately
after the change of the value of $q$ the differences between the domains in the perturbed and in the control systems
strongly depend on when the change took place,
which gives raise to an increase of $M(t,s)$ with the waiting time $s$. However, once the ordered domains in the
perturbed system have undergone major changes, they become similar to those that are formed in the control system,
which results in a weaker dependence on $L_x(s)$.

\section{Conclusion}
Phase-ordering kinetics in systems with algebraic domain growth, especially in cases where the domain coarsening
is curvature driven, is rather well understood. Still open problems are encountered in disordered systems where after
a transient algebraic-like growth regime a non-algebraic, logarithmic domain growth prevails for long times.

In this paper we discussed some aspects of domain coarsening in systems of a different type that are also
characterized  by logarithmic growth, namely in systems dominated by dynamical constraints. The best known
representative of this class is the $ABC$ model \cite{Eva98a,Eva98b} where in a particular direction, say
the horizontal direction, particles of three different types 
are exchanged asymmetrically. This leads to the formation of ordered domains which in a two-dimensional system take
the form of stripes oriented perpendicularly to the direction of biased exchanges \cite{Kaf00}. In the asymptotic long-time regime,
these domains coarsen and their size increases logarithmically with time.

Our emphasis was on the ordering process and the related aging phenomena. Studying a time-dependent length derived
from the single-time spatial correlator, we identified different regimes that we discussed as a function of 
the vertical extent of the system and of the bias in the particle exchanges. Whereas in the logarithmic growth regime
dynamical scaling prevails, as demonstrated  by the data collapse of the single-time correlator as well as of the
two-times autocorrelation (and also, to some extent, of the two-times response function), complicated crossover
phenomena are observed during the formation of the well ordered stripes, especially in vertically extended systems
where an intermediate phase is observed that is dominated by the vertical growth of the domains. 

The attentive reader will have noticed that in the discussion of our results we never brought up the fact that for
the one-dimensional system with even coverage for all three species the steady state is an equilibrium steady state,
whereas in the two-dimensional version of the model detailed balance is always violated and the steady state
is a non-equilibrium steady state. In fact, the equilibrium or non-equilibrium nature of the steady state does not seem to be relevant 
for the dynamical, far-from-stationarity regime probed in our study. While it seems reasonable
that a system in the stages of domain formation and subsequent domain coarsening does not know anything about
the character of the steady state emerging in the long time limit (see \cite{Oli93,Dro99,And06}
for another example where the dynamic properties during relaxation do not depend on the equilibrium or
non-equilibrium nature of the stationary state), there do exist well known examples (many-particle systems
without detailed balance \cite{Hen07}, driven Ising systems \cite{Cor96,God11,God14}, driven striped structures \cite{Eva00}) 
where for cases with a non-equilibrium steady state
dynamic properties during the relaxation process are different from those encountered in the corresponding systems with an
equilibrium steady state. Additional
studies will be needed in the future in order to fully comprehend these differences.

\begin{acknowledgments}

This research is supported by the U.S. Department of Energy,
  Office of Basic Energy Sciences, Division of Materials Sciences and
  Engineering under Award DE-FG02-09ER46613.

\end{acknowledgments}


\begin{thebibliography}{99}

\bibitem{Cro93} M. C. Cross and P. C. Hohenberg, Rev. Mod. Phys. {\bf 65}, 851 (1993).

\bibitem{Bra94} A. J. Bray, Adv. Phys. {\bf 43}, 357 (1994).

\bibitem{Hen10} M. Henkel and M. Pleimling, {\it Non-Equilibrium Phase Transitions,
Vol. 2: Ageing and Dynamical Scaling Far from Equilibrium}
(Springer, Berlin, 2010).

\bibitem{Cat15} M. E. Cates, in {\it Soft Interfaces}, edited by D. Quer\'{e},
L. Bocquet, T. Witten, and L. F. Cugliandolo, Les Houches XCVIII (Oxford University Press,
2015); arXiv:1209.2290.

\bibitem{Szo14} A. Szolnoki, M. Mobilia, L.-L. Jiang, B. Szczesny, A. M. Rucklidge, and M. Perc,
J. Roy. Soc. Interface {\bf 11}, 0735 (2014)

\bibitem{Rom13} A. Roman, D. Dasgupta, and M. Pleimling, Phys. Rev. E {\bf 87}, 032148 (2013).

\bibitem{Cas09} C. Castellano, S. Fortunato, and V. Loreto, Rev. Mod. Phys. {\bf 81}, 591 (2009).

\bibitem{Pur09} S. Puri and V. Wadhawan (editors), {\it Kinetics of phase transitions}
(CRC Press, Boca Raton, 2009).

\bibitem{Voo85} P. W. Voorhees, J. Stat. Phys. {\bf 38}, 231 (1985).

\bibitem{Cug14} L. F. Cugliandolo, arXiv:1412.0855.

\bibitem{Kol05} A. B. Kolton, A. Rosso, and T. Giamarchi, Phys. Rev. Lett. {\bf 95}, 180604 (2005).

\bibitem{Noh09} J. D. Noh and H. Park, Phys. Rev. E {\bf 80}, 040102(R) (2009).

\bibitem{Igu09} J. L. Iguain, S. Bustingorry, A. B. Kolton, and L. F. Cugliandolo, Phys. Rev. B
{\bf 80}, 094201 (2009).

\bibitem{Mon09} C. Monthus and T. Garel, J. Stat. Mech.: Theory Exp. (2009) P12017.

\bibitem{Rao93} M. Rao and A. Chakrabarti, Phys. Rev. Lett. {\bf 71}, 3501 (1993).

\bibitem{Aro08} C. Aron, C. Chamon, L. F. Cugliandolo, and M. Picco, J. Stat. Mech.: Theory Exp. (2008) P05016.

\bibitem{Par10} H. Park and M. Pleimling, Phys. Rev. B {\bf 82}, 144406 (2010).

\bibitem{Cor11} F. Corberi, E. Lippiello, A. Mukherjee, S. Puri, and M. Zannetti, 
J. Stat. Mech.: Theory Exp. (2011) P03016.

\bibitem{Cor12} F. Corberi, E. Lippiello, A. Mukherjee, S. Puri, and M. Zannetti,
Phys. Rev. E {\bf 85}, 021141 (2012).

\bibitem{Par12} H. Park and M. Pleimling, Eur. Phys. J. B {\bf 55}, 300 (2012).

\bibitem{Cor13} F. Corberi, E. Lippiello, A. Mukherjee, S. Puri, and M. Zannetti,
Phys. Rev. E {\bf 88}, 042129 (2013).

\bibitem{Man14} P. K. Mandal and S. Sinha, Phys. Rev. E {\bf 89}, 042144 (2014).

\bibitem{Fis01} D. S. Fisher, P. Le Doussal, and C. Monthus, Phys. Rev. E {\bf 64}, 066107 (2001).

\bibitem{Cor02} F. Corberi, A. de Candia, E. Lippiello, and M. Zannetti,
Phys. Rev. E {\bf 65}, 046114 (2002).

\bibitem{Eva02} M. R. Evans, J. Phys.: Condens. Matter {\bf 14}, 1397 (2002).

\bibitem{Eva98a} M. R. Evans, Y. Kafri, H. M. Koduvely, and D. Mukamel, Phys. Rev. Lett. {\bf 80}, 425 (1998).

\bibitem{Eva98b} M. R. Evans, Y. Kafri, H. M. Koduvely, and D. Mukamel, Phys. Rev. E {\bf 58}, 2764 (1998).

\bibitem{Lah97} R. Lahiri and S. Ramaswamy, Phys. Rev. Lett. {\bf 79}, 1150 (1997).

\bibitem{Lah00} R. Lahiri, M. Barma, and S. Ramaswamy, Phys. Rev. E {\bf 61}, 1648 (2000).

\bibitem{Lip09} A. Lipowski and D. Lipowska, Phys. Rev. E {\bf 79}, 060102(R) (2009).

\bibitem{Cli03} M. Clincy, B. Derrida, and M. R. Evans, Phys. Rev. E {\bf 67}, 066115 (2003).

\bibitem{Bod08} T. Bodineau, B. Derrida, V. Lecomte, and F. van Wijland, J. Stat. Phys. {\bf 133}, 1013 (2008).

\bibitem{Ayy09} A. Ayyer, E. A. Carlsen, J. L. Lebowitz, P. K. Mohanty, D. Mukamel, and E. Speer,
J. Stat. Phys. {\bf 137}, 1166 (2009).

\bibitem{Led10a} A. Lederhendler and D. Mukamel, Phys. Rev. Lett. {\bf 105}, 150602 (2010).

\bibitem{Led10b} A. Lederhendler, O. Cohen, and D. Mukamel, J. Stat. Mech.: Theory Exp. (2010) P11016.

\bibitem{Bar11a} J. Barton, J. L. Lebowitz, and E. R. Speer, J. Phys. A: Math. Theor. {\bf 44}, 065005 (2011).

\bibitem{Ber11} L. Bertini, N. Cancrini, and G. Posta, J. Stat. Phys. {\bf 144}, 1284 (2011).

\bibitem{Bar11b} J. Barton, J. L. Lebowitz, and E. R. Speer, J. Stat. Phys. {\bf 145}, 763 (2011).

\bibitem{Bod11} T. Bodineau and B. Derrida, J. Stat. Phys. {\bf 145}, 745 (2011).

\bibitem{Coh11} O. Cohen and D. Mukamel, J. Phys. A: Math. Theor. {\bf 44}, 415004 (2011).

\bibitem{Ger11} A. Gerschenfeld and B. Derrida, Europhys. Lett. {\bf 96}, 20001 (2011).

\bibitem{Coh12a} O. Cohen and D. Mukamel, Phys. Rev. Lett. {\bf 108}, 060602 (2012).

\bibitem{Ger12} A. Gerschenfeld and B. Derrida, J. Phys. A: Math. Theor. {\bf 45}, 055002 (2012).

\bibitem{Coh12b} O. Cohen and D. Mukamel, J. Stat. Mech.: Theory Exp. (2012) P12017.

\bibitem{Afz13} N. Afzal and M. Pleimling, Phys. Rev. E {\bf 87}, 012114 (2013).

\bibitem{Ber13} L. Bertini and P. Butt\`{a}, J. Stat. Phys. {\bf 152}, 15 (2013).

\bibitem{Coh14} O. Cohen and D. Mukamel, Phys. Rev. E {\bf 90}, 012107 (2014).

\bibitem{Mis14} R. Misturini, arXiv:1403.4981.

\bibitem{Kaf00} Y. Kafri, D. Biron, M. R. Evans, and D. Mukamel, Eur. Phys. J. B {\bf 16}, 
669 (2000).

\bibitem{Bra93} A. J. Bray, Physica A {\bf 194}, 41 (1993).

\bibitem{Oli93} M. J. de Oliveira, J. F. F. Mendes and M.A. Santos,
J. Phys. A: Math. Gen. {\bf 26}, 2317 (1993).

\bibitem{Dro99} J. M. Drouffe and C. Godr\`{e}che, J. Phys. A: Math. Gen. {\bf 32}, 249 (1999).

\bibitem{And06} N. Andrenacci, F. Corberi, E. Lippiello, Phys. Rev. E {\bf 73}, 046124 (2006).

\bibitem{Hen07} M. Henkel, J. Phys. Condens. Matter {\bf 19}, 065101 (2007).

\bibitem{Cor96} S. J. Cornell and A. J. Bray, Phys. Rev. E {\bf 54}, 1153 (1996).

\bibitem{God11} C. Godr\`{e}che, J. Stat. Mech.: Theory Exp. (2011) P04005.

\bibitem{God14} C. Godr\`{e}che and M. Pleimling, J. Stat. Mech.: Theory Exp. (2014)  P05005.

\bibitem{Eva00} M. R. Evans, Y. Kafri, E. Levine, and D. Mukamel, Phys. Rev. E {\bf 62}, 7619 (2000).

\end{thebibliography}
\end{document}